# High-pressure synthesis of pure and doped superconducting MgB$_2$ compounds

**P. Toulemonde,**

N. Musolino and R. Flükiger.

*Département de Physique de la Matière Condensée, Université de Genève,*

*24 quai Ernest-Ansermat, CH-1211 Genève 4, Switzerland.*

**Abstract**

Superconducting properties of bulk, dense, pure $MgB_2$ and doped $(Mg_{1-x}A_x)B_2$ samples with A = Na, Ca, Cu, Ag, Zn and Al were studied for compositional ranges $0 < x ? 0.20$. The effect on pinning properties and critical current were investigated, particularly for A = aluminium. The samples were sintered and/or synthesised at high pressure – high temperature in a cubic multi-anvils press (typically : 3.5-6 GPa, 900-1000 °C). They were characterised by x-ray diffraction, scanning electron microscopy and their superconducting properties were investigated by a.c. susceptibility, magnetisation (VSM and SQUID) and transport measurements under magnetic field. Only Al substitutes really on the Mg site. The other elements form secondary phases with B or Mg which do not act as pinning centres. No positive effect is observed on the superconducting properties of the bulk $MgB_2$ samples added with these doping elements : $T_c$, $j_c$ critical current, $H_{irr}$ and $H_{c2}$. For Al, the effect on $H_{c2}$ remains small, and the irreversibility line does not move, thus not improving the critical current of the Al-doped $MgB_2$ samples.

**Introduction**

Since the discovery of superconductivity in $MgB_2$ at 39 K [1], this material has been prepared under a variety of conditions and its quality is sample-dependant. Due to the low thermal stability of $MgB_2$, its preparation has to be done in closed systems [2, 3]. The most common method is based on the reaction between a fine powder of boron and liquid magnesium in equilibrium with its vapour above the melting point of Mg (650 °C) in a sealed metallic tube of Ta or Mo. This technique at low pressure is currently used to prepare $MgB_2$ pellets [4] (or wires by diffusion of Mg into B fibers [5]), but the polycrystalline samples obtained are porous and not suitable for good transport measurements. A second method, based on high pressure process, allows one to synthesise dense $MgB_2$ samples. For over 40 years, magnesium has been used as a catalyst for the transformation of hexagonal boron nitride (BN) to its cubic form at high pressure – high temperature (HP-HT) [6-8] ; during this HP-HT treatment the formation of secondary $MgB_6$ and $MgB_2$ phases was observed [9]. Recently, the HP-HT method has been applied to $MgB_2$ for its sintering [10-12], its synthesis [10, 13-15], and its single crystal growth [16-18]. This paper deals with the HP-HT sintering of $MgB_2$, its HP-HT synthesis and the HP-HT preparation of doped $(Mg_{1-x}A_x)B_2$ compounds with A = Al, Zn, Cu, Ag, Na, Ca. The superconducting properties of all the samples will be discussed. The attempts to substitute Mg for another element had two main purposes : firstly to find a way to increase $T_c$ in this system and secondly to introduce defects in the $MgB_2$ structure in order to decrease the mean free path of the normal electrons decrease the coherence length ? and consequently increase $H_{c2}$ and also to introduce local pinning centres to increase the irreversibility line. Indeed, the position of this line is quite low in $MgB_2$ ($H_{irr}$ (0 K) ? 10 T) and well below $H_{c2}$ (? 18 T at 0 K, see ref. [10]) compared to those of the high $T_c$ cuprates.

1. **HP-HT sintered pure MgB$_2$ samples**

HP-HT sintered MgB$_2$ samples were first prepared by Y. Takano et al. [11] and C.U. Jung et al. [12]. Both worked from a commercial MgB$_2$ powder. The HP-HT sintering consists of preparing a pellet at ambient pressure, and inserting it into a BN (or Au) crucible which is mounted in the centre of a pyrophillite high pressure cell. The whole assembly is placed between the anvils of the press and the pressure is increased up to a constant level. A tubular carbon furnace surrounds the sample, and allows the MgB$_2$ pellet to be heated up to a plateau which varies from few minutes to several hours. Both groups tried different conditions of pressure, temperature and time and have obtained the best superconducting properties for P ? 3-3.5 GPa and T ? 900-950 °C. These were also the conclusion of V.N. Narozhnyi very recently [19]. Using the same technique than Y. Takano and C.U. Jung, one year ago, we sintered dense and high quality MgB$_2$ samples, starting from a commercial MgB$_2$ powder (Alfa, 98 %). Our samples were suitable for critical current measurements by magnetisation and transport techniques. We have showed that the critical current in dense polycrystalline MgB$_2$ samples is not affected by weak links at the grain boundaries [10], like in cuprates.

2. **HP-HT synthesised pure MgB$_2$ samples**

Shortly after the discovery of superconductivity in MgB$_2$, we were the first to synthesise MgB$_2$ in HP-HT conditions, starting directly from metallic Mg and B powders [10]. Since then, other groups have prepared similar samples [13-15]. The method consists of a reaction between liquid magnesium and solid boron at HP-HT. It is similar to the method of HP-HT sintering. The powders were mixed together in a stoichiometric ratio, inserted into a BN crucible and placed in the cubic high pressure cell made of pyrophyllite. The same cubic multi-anvils press was used to apply a pressure of 3.5-4.5 GPa and the temperature was maintained at 850-950 °C for 1-2 h during the HP-HT synthesis. The conditions (pressure, time and temperature) were optimised to obtain the sharpest superconducting transition.

These samples were prepared from Mg and B of different granulometry : flakes or fine powder (Alfa, 99.8%, 325mesh) for Mg and crystalline powder for B (Ventron, m2N7, 60 mesh and Alfa, 99.7%, 325 mesh). They were characterised by x-ray diffraction using copper $K_{???}$ radiation (? = 1.544 Å) ; the XRD patterns of three different samples are presented in figure 1. The main impurity is MgO, whose concentration increases with the powder size of the Mg starting material. Combining fine crystalline Mg and B powders gives the purest sample (A). The lattice parameters of the HP-HT synthesised $MgB_2$ samples, calculated by a least square method from the reflections list, are consistent with the literature [20] : a = 3.083(4) Å, c = 3.518(2) Å for sample A. The microstructure of sample A is shown in figure 2 ; it shows small, well shaped $MgB_2$ grains of 1-10 µm (similar to those of the HT-HT synthesised sample in ref. [15]) and small MgO particles. Back scattered electron images show evidence of boron rich dark regions (around 10 µm) corresponding to not yet fully reacted boron grains. Their amorphous nature is indicated by XRD measurements. The presence of these regions explain why the apparent $j_c$ value is lower for synthesised samples than for sintered samples from a commercial $MgB_2$ powder (around $10^6$ $A/cm^2$ at 4.2 K in zero field, figure 3) ; indeed, the whole volume of the sample (including the non superconducting boron rich regions) is taken into account in the $j_c$ calculations from the magnetisation data obtained in a VSM and analysed using the Bean model (see ref. [10] for details).

The figure 4 shows the a.c. susceptibility of the three samples. The best result is obtained for the mixture of very fine reacting powders (325 mesh ~ 40 µm). The superconducting transition is very broad (around 10 K) for the sample prepared from Mg flakes. This width decreases as the size of the reacting powders decreases, reaching ~ 1.5 K for the best sample (A). The $j_c$ critical current measured for this sample by magnetisation (VSM) was estimated to be 2?$10^5$ A/cm2 at 4.2 K in zero field and $10^4$ A/cm$^2$ at 20 K, in a magnetic field of 2 T (figure 3).

### 3. HP-HT synthesised substituted ($Mg_{1-x}A_x$)$B_2$ samples

Initially, different substitutions on the Mg site were tested experimentally [21-33] and theoretically [34-36] in an attempt to increase the $T_c$ of $MgB_2$. Additive elements were also investigated to increase the pinning potential of the material in order to reach higher $j_c$ values. Except for aluminium [21], few elements enter the $MgB_2$ lattice and substitute on the Mg site to give a real solid solution ($Mg_{1-x}A_x$)$B_2$.

#### 3.1. HP-HT synthesis

Most syntheses of doped $MgB_2$ materials have been made at ambient pressure ; this limits the number of possible substitutions. The HP-HT process may increase the solubility limit of the element in substitution for Mg. Here, we have tried different substitutions for A = Al, Zn, Cu, Ag, Na, Ca with x = 0.1 in our multi-anvils press. Other compositions for 0 < x < 0.20 were studied for Zn and Al (see paragraph 3.3). The additive element was mixed with the Mg and B fine powders in a stoichiometric ratio and the synthesis conditions were kept close to those for pure $MgB_2$ samples (table 1).

#### 3.2. Characterisation

Figure 5 shows the XRD patterns of $MgB_2$ doped samples with x = 0.10 of Na, Ca, Cu, Zn and Ag. The main phase is still $MgB_2$, but some impurities appear : NaCl, $CaB_6$, $Cu_2Mg$, Zn + $MgZn_2$ and MgAg respectively. The lattice parameters of the $MgB_2$ phase in these doped samples do not change significantly and stay very close to the initial values for undoped materials (table 1.). This means that these elements do not substitute for Mg. This has been confirmed by C.H. Cheng et al. who observed very recently that the solubility limit of Ag at Mg site is around x = 0.005 [37]. For Zn, and contrary to Kazakov et al. [22], and even using such severe conditions like thus of HP-HT synthesis, we found no evidence for a change in the a and c lattice parameters in our samples (x = 0.05, 0.10 and 0.20), within experimental errors. If Zn really occupies the Mg site, its solubility limit is low.

SEM images (from back electrons, figure 6) confirm the presence of the secondary phases detected by XRD. The typical size of these impurity grains varies from 1 μm (small rounded grains of MgO) to 20 μm (for $Cu_2Mg$ or $MgZn_2$ for instance).

A.C. susceptibility measurements show that these doped samples have a slightly lower $T_c$, possibly due to under-stoichiometry on Mg site. Only the sample doped with 5 % of Zn shows a slightly higher $T_c$ (inductive) than others samples (table 1). This is confirmed by resistivity measurements, as shown in figure 7 : $T_c$ onset for the 5 % Zn doped sample is 0.3 K higher than for the pure $MgB_2$ synthesised in similar HP-HT conditions (respectively 38.9 and 38.6 K). Nevertheless, considering XRD and EDX data, it is not possible to conclude that this slight increase is related to Zn substitution for Mg site. For Zn doping, the samples show a continuous broadening of the superconducting transition with increasing concentration (table 1 and figure 8 for Zn). This broadening exists also for Al, but is less marked.

The defects made up of the grains of secondary phases (for Na, Ca, Cu, Zn and Ag) could act as pinning centres and may increase the critical current. However, the magnetisation measurements (VSM) show that these inclusions do not really improve $j_c$ nor its magnetic field dependence. The curves displayed in figure 9 for 5 % Zn doped, 10 % Na doped and pure MgB$_2$ are close to each other. The size of the impurity grains (> 1 μm) is greater than the typical coherence length of MgB$_2$, so they can not be efficient pinning centres. Apart from a slightly better j$_c$ field dependence in 5 % Zn doped sample, the interest of such doping elements is then limited.

### 3.3. Aluminium doping

The first study of Al doped MgB$_2$ has shown a slight decrease in T$_c$ for x ? 0.1 [21] ; for 0.1 ? x ? 0.25, a two phase mixture is found, whereas for x > 0.25 no bulk superconductivity is detected. Furthermore, for x > 0.025, a splitting of the (002) reflection has been observed by Pissas et al. [38] by synchrotron XRD, which shows the existence of two different phases. However, A. Bianconi et al. has observed superconductivity up to x = 0.5 at 4 K in (Mg$_{1-x}$Al$_x$)B$_2$ [39, 40].

Our study was restricted to (Mg$_{1-x}$Al$_x$)B$_2$ samples with small Al contents : x = 0.01, 0.03, 0.05, 0.10 and 0.20. Here, the purpose was to study the effects on pinning and j$_c$ field dependence of defects and disorder induced in the Mg plane of MgB$_2$ by Al atoms substituted for Mg atoms. Others aspects of the Al substitution have been studied in the literature [33, 41-48], but not the critical behaviour.

The samples were synthesised using the same procedure described above, at high pressure – high temperature (see table 1.). As expected [21], the lattice parameters of our Al doped samples decrease with Al content : 0.21 Å/x for a-axis and four times faster for c-axis : 0.88 Å/x. A gradual decrease of T$_{c\,onset}$ from 38 K (x = 0.01) to 36 K (x = 0.10) with increasing Al content is also measured below x = 0.10 (in the single-phase region [21]), but the superconducting transitions (from SQUID measurements, figure 10) do not broaden as much as in ref. [21] (table 1). This suggests that the use of HP-HT process allows us to substitute Al for Mg more homogeneously in our samples. This technique may yield single phase samples with x > 0.10 in the future.

The effect of Al substitution in the Mg planes on the critical current is visible in figure 9 (see curve of 5% Al doped sample). For x = 0.05 of the solid solution, the $j_c$ behaviour as a function of magnetic field is degraded, compared to the pure $MgB_2$ sample. Then, the defects constituted by these Al atoms in the Mg plane do not play the role of efficient pinning centres.

In these Al doped $MgB_2$ compounds, where a continuous solid solution exist up to x = 0.10, the local defects introduced by Al on the Mg site should decrease the mean free path "l" of the charge carriers and thus increase the resitivity ? (in the normal state), and consequently increase $H_{c2}$ which is proportional to $?_0/(?_0 l)$. For x = 0.05, resistivity measurements were performed under different magnetic fields up to 14 T and the upper critical field $H_{c2}$ was extracted from the onset of the superconducting transition and the irreversibility field $H_{irr}$ from the offset. They were compared to $H_{c2}$ and $H_{irr}$ determined respectively from the shift in $T_c$ in the Meissner regime measured under different fields and magnetisation loops in a SQUID magnetometer (figure 11). No significant increase of $H_{irr}$ is observed (extrapolated value : $H_{irr}$ (0K) ~ 10 T), compared to the value for pure $MgB_2$ [10], but a slight increase of extrapolated $H_{c2}$ (0 K) from ~ 18 T (x = 0, [10]) to ~ 21 T for x = 0.05 was obtained (figure 12). This result is yet to be confirmed for x ? 0.05.

**Conclusion**

We have sintered and synthesised pure bulk $MgB_2$ and doped $(Mg_{1-x}A_x)B_2$ compounds (A = Na, Ca, Cu, Zn, Ag and Al for x ? 0.20) at high temperature - high pressure using a cubic multi-anvils press. This HP-HT process allows us to obtain dense polycrystalline samples suitable for transport measurements. Zn and Al doping were more carefully investigated. XRD and SEM studies show that only Al is really substituted for Mg. The other doping elements react with Mg or B to form secondary phases. The $T_{c\ onset}$ value of these doped $MgB_2$ materials is largely unaltered, except in the case of Al where it decreases continuously with the Al content. The most important effect is a broadening of the superconducting transition. The magnetisation measurements, using VSM and SQUID, show that $j_c$ is not improved and $H_{c2}$ is only slightly increased for the Al substituted samples. The impurity inclusions (non superconducting) are not efficient pinning centres, because their size is too large compared to the coherence length of $MgB_2$. The defects constituted by the random Al distribution in the Mg plane do not play this role either. This work suggests that the origin of increased critical currents reported in bulk samples with non sub-micron grain size is perhaps due to defect pinning centres (oxygen, carbon) in the boron plane of the $MgB_2$ grains. The research and the control of such intra-grains nano-defects is underway.

**Acknowledgment**

P. Toulemonde would like to thank A. Naula for his help in the technical preparation of the HP-HT experiments performed in the multi-anvils press and Dr. C. Beneduce for the resistive measurements.

**Figure captions**

**Figure 1.** X-ray diffraction pattern of high pressure – high temperature synthesised $MgB_2$ samples with Mg and B of different granulometry at ? = 1.544 Å.

**Figure 2.** SEM image of HP-HT synthesised $MgB_2$ with crystalline fine Mg and B powders (325mesh).

**Figure 3.** Inductively measured $j_c$ values plotted against applied magnetic field for different temperatures (10, 20 and 30 K) in HP-HT sintered and HP-HT synthesised $MgB_2$ samples.

**Figure 4.** A.C. susceptibility measurements of the superconducting transition for the HP-HT synthesised $MgB_2$ samples.

**Figure 5.** X-ray diffraction pattern at ? = 1.5418 Å of HP - HT synthesised $(Mg_{1-x}A_x)B_2$ samples with A = Na, Ca, Cu, Zn and Ag, x = 0.10.

**Figure 6.** Electron back scattered image of HP-HT synthesised $(Mg_{0.90}Zn_{0.10})B_2$ sample.

**Figure 7.** D.C. resistive measurement of superconducting transitions in HP-HT sintered and synthesised pure and 5 % Zn doped $MgB_2$ samples. $T_{c\ onset}$ and $T_{c0}$ are indicated.

**Figure 8.** A.C. susceptibility measurements of the superconducting transition for the HP-HT synthesised doped $(Mg_{1-x}Zn_x)B_2$ samples with x = 0.05, 0.10 and 0.20.

**Figure 9.** Inductively measured $j_c$ values plotted against applied magnetic field for different temperatures (10, 20 and 30 K) in 5 % Zn, 5 %Al, 10 % Na doped and pure $MgB_2$ samples.

**Figure 10.** Zero field cooled and field cooled magnetisation of $(Mg_{1-x}Al_x)B_2$ as a function of temperature and Al concentration in a field of 20 Oe.

**Figure 11.** Magnetisation of $(Mg_{0.95}Al_{0.05})B_2$ sample at different temperature and fields up to 5 T.

**Figure 12.** The temperature dependence of the second critical field $H_{c2}$ and of the irreversibility field $H_{irr}$ measured inductively (VSM or SQUID data)and resistively for $(Mg_{0.95}Al_{0.05})B_2$ sample.

**Tables**

**Table 1.** HP-HT synthesis conditions, inductive $T_c$, and lattice parameters of the $(Mg_{1-x}A_x)B_2$ samples with A = Na, Al, Ca, Cu, Zn and Ag for $0 < x ? 0.20$.

| doping element | content x | synthesis conditions | $T_c^{onset}$ (K) | $\Delta T_c$ (K) | lattice parameters a (Å) | lattice parameters c (Å) |
|---|---|---|---|---|---|---|
| **Na** (NaCl) | 0.10 | 4 GPa, 950°C, 1h | 38 | 3 | 3.082(5) | 3.517(8) |
| **Al** | 0.01 | 3.5 GPa, 950°C, 1h | 38 | 3.5 | 3.084(1) | 3.517(1) |
| | 0.03 | 3.5 GPa, 950°C, 1h | 38 | 3.5 | 3.081(1) | 3.511(1) |
| | 0.05 | 3.5 GPa, 950°C, 1h | 37 | 4.5 | 3.078(2) | 3.492(2) |
| | 0.10 | 3.5 GPa, 900°C, 1h | 36 | 6 | 3.065(5) | 3.437(2) |
| **Ca** (CaH$_2$) | 0.10 | 3.5 GPa, 900°C, 1h | 38 | 3 | 3.084(2) | 3.526(1) |
| **Cu** | 0.10 | 4 GPa, 900°C, 1h | 38.5 | 2 | 3.076(4) | 3.519(3) |
| **Zn** | 0.05 | 3.5 GPa, 950°C, 1h | 39 | 4 | 3.085(2) | 3.523(2) |
| | 0.10 | 3.5 GPa, 950°C, 1h | 38 | 2.5 | 3.083(1) | 3.523(1) |
| | 0.20 | 6 GPa, 900°C, 1h | 38 | 4 | 3.082(8) | 3.513(7) |
| **Ag** | 0.10 | 3.5 GPa, 1000°C, 1h | 38.5 | 2.5 | 3.080(1) | 3.518(1) |

Table 1.

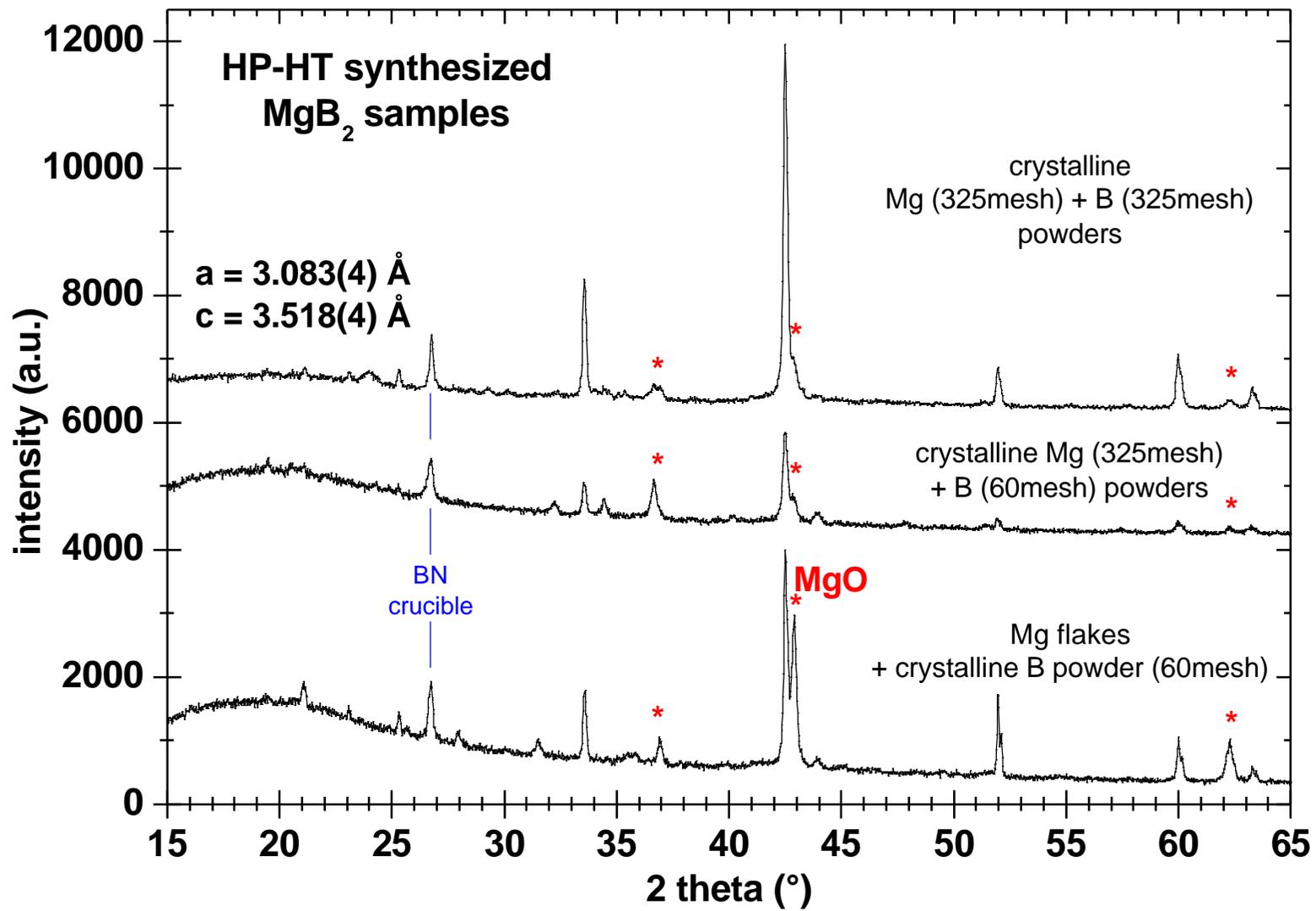

Fig. 1.

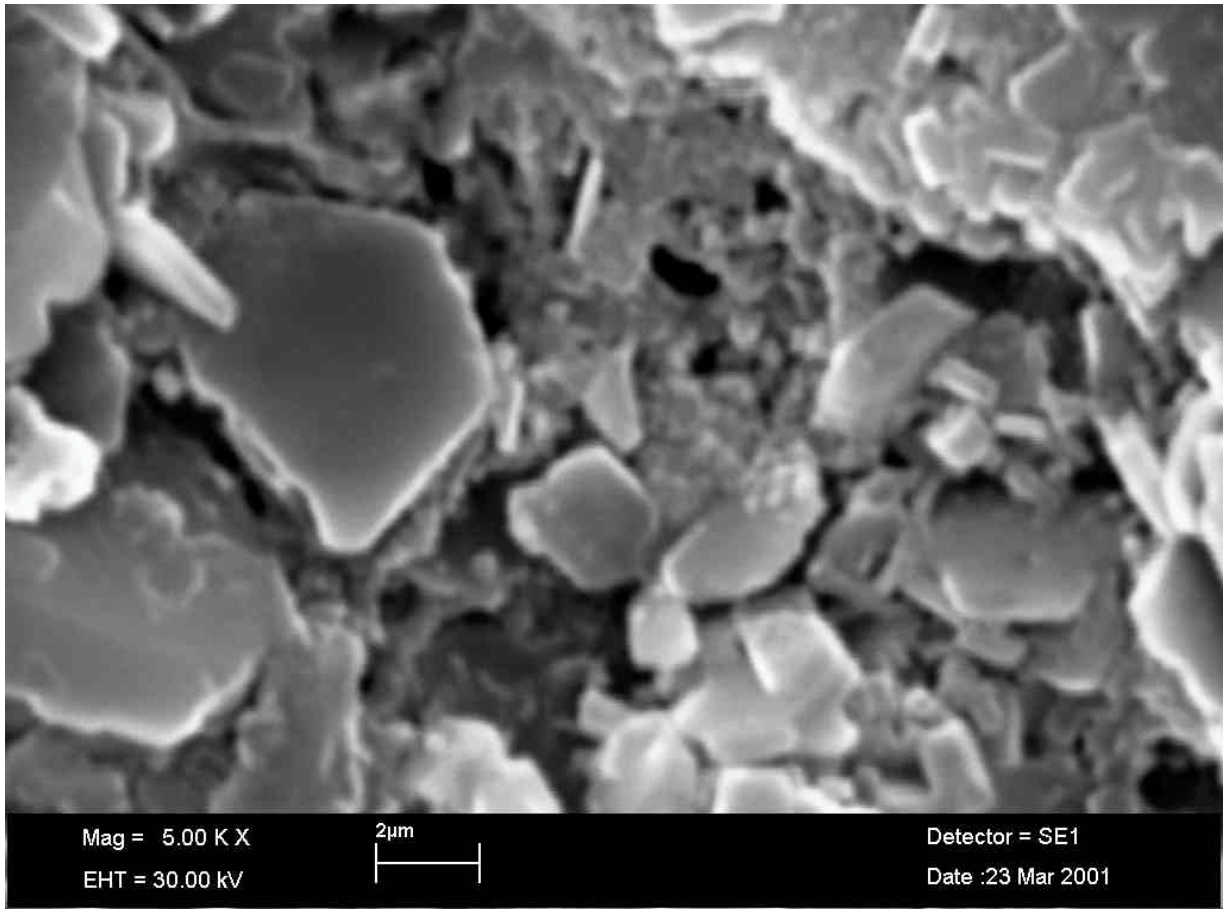

Fig. 2.

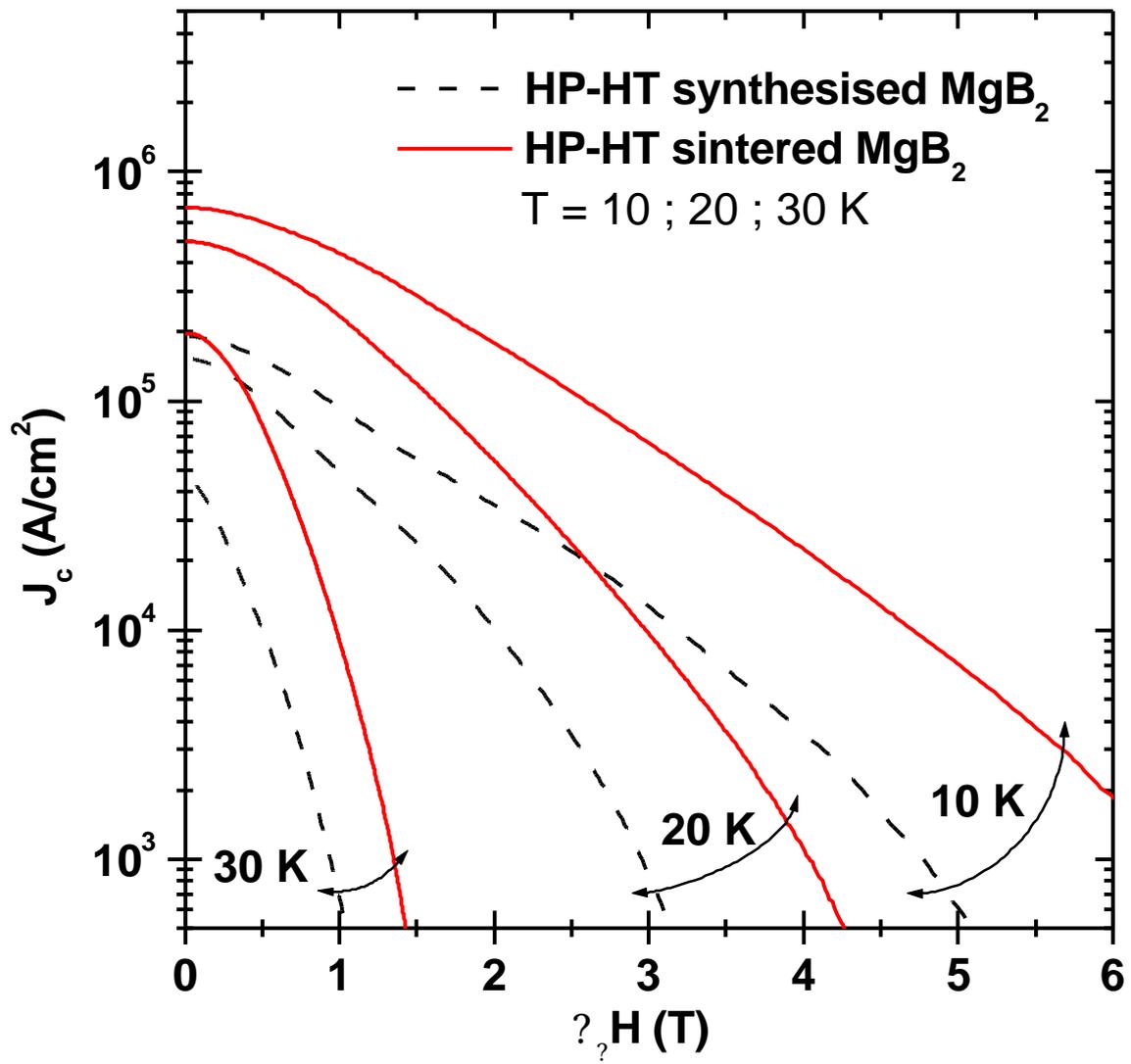

Fig. 3.

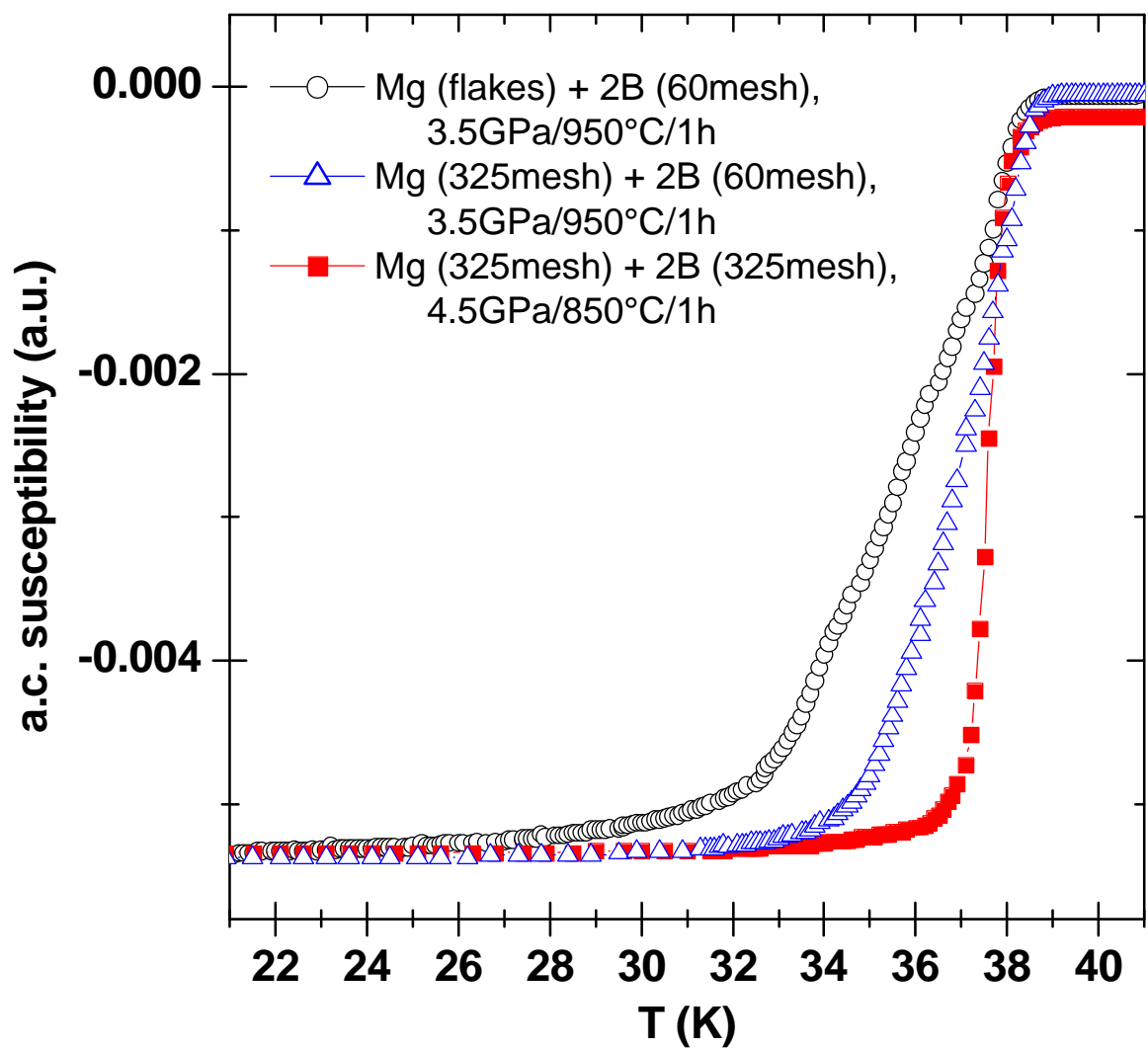

Fig. 4.

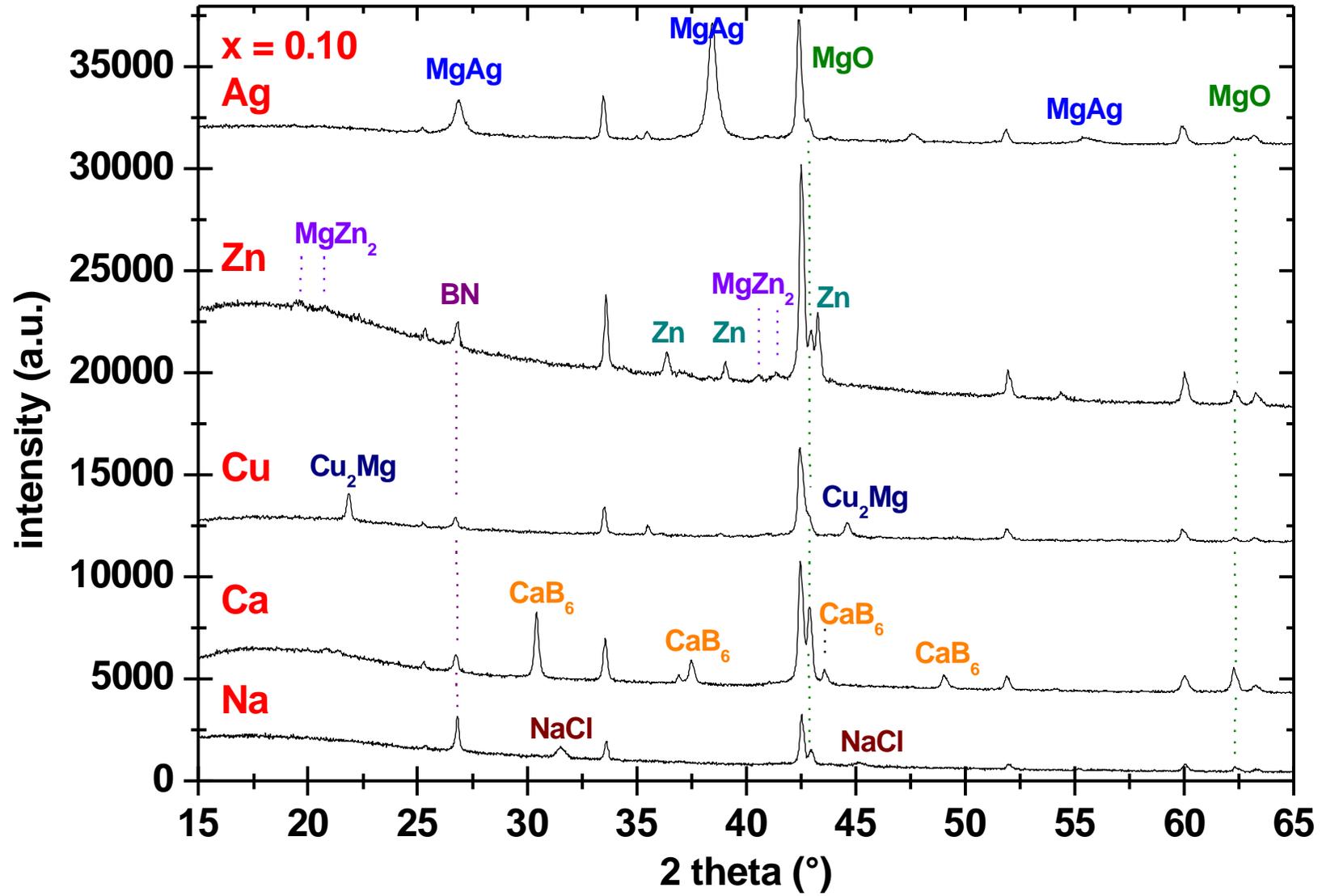

Fig. 5.

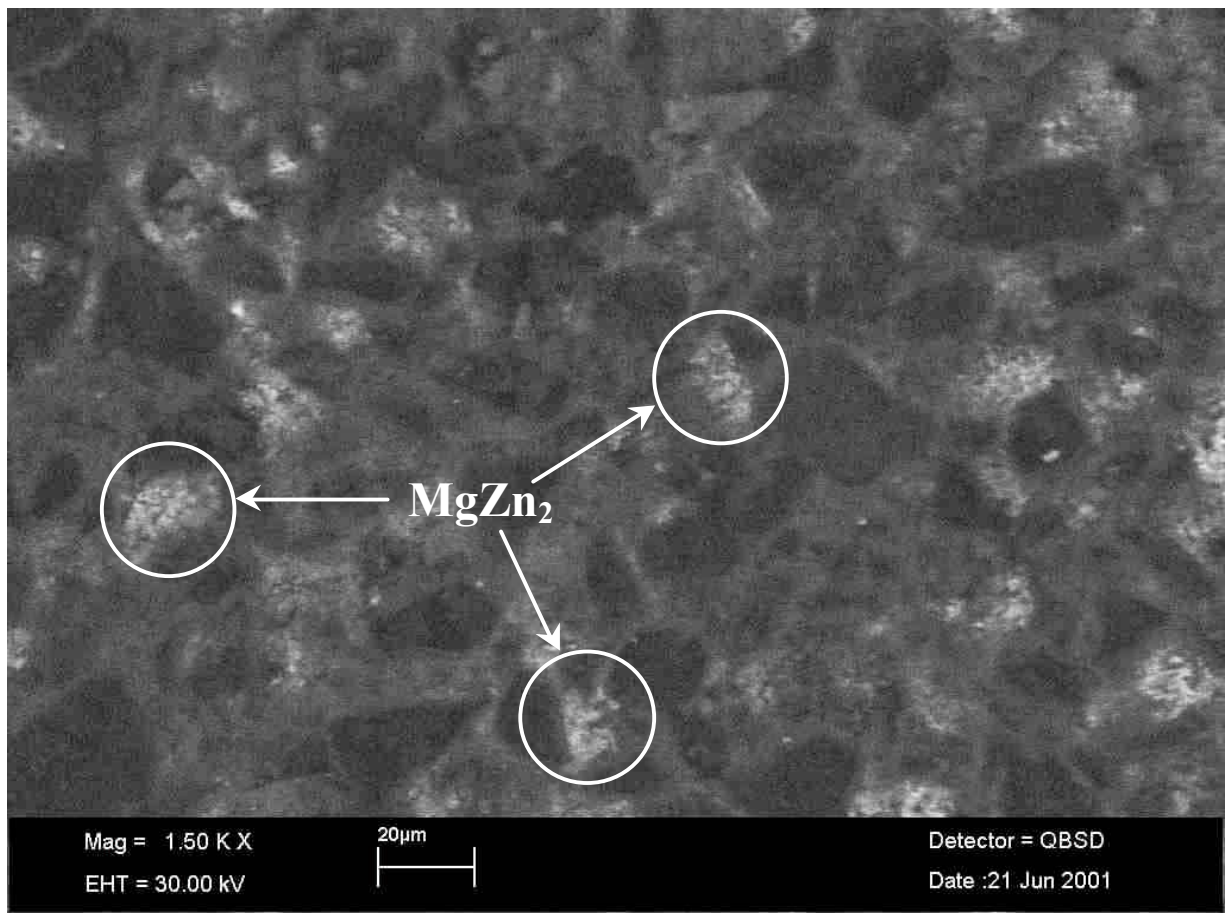

Fig. 6.

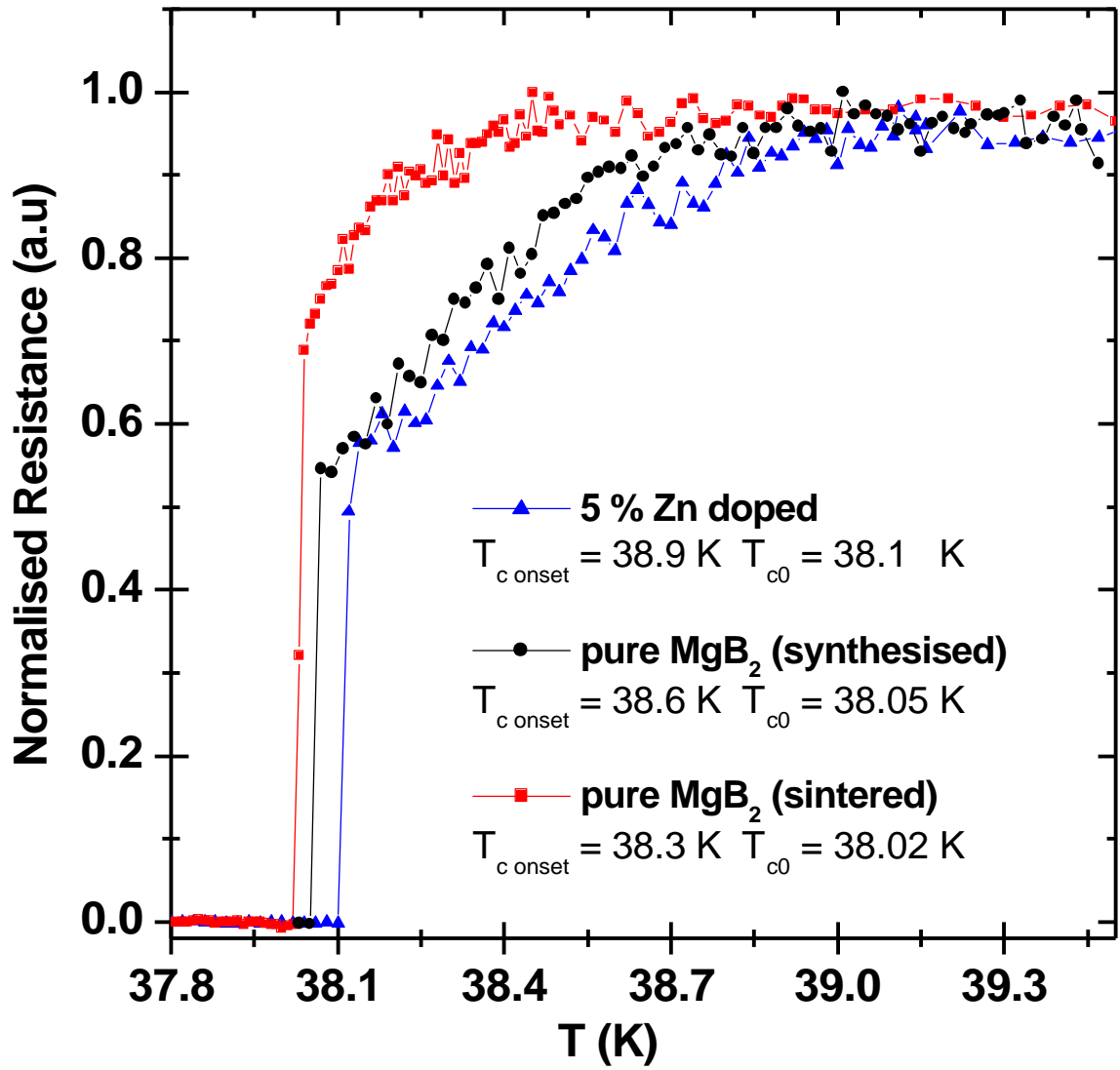

Fig. 7.

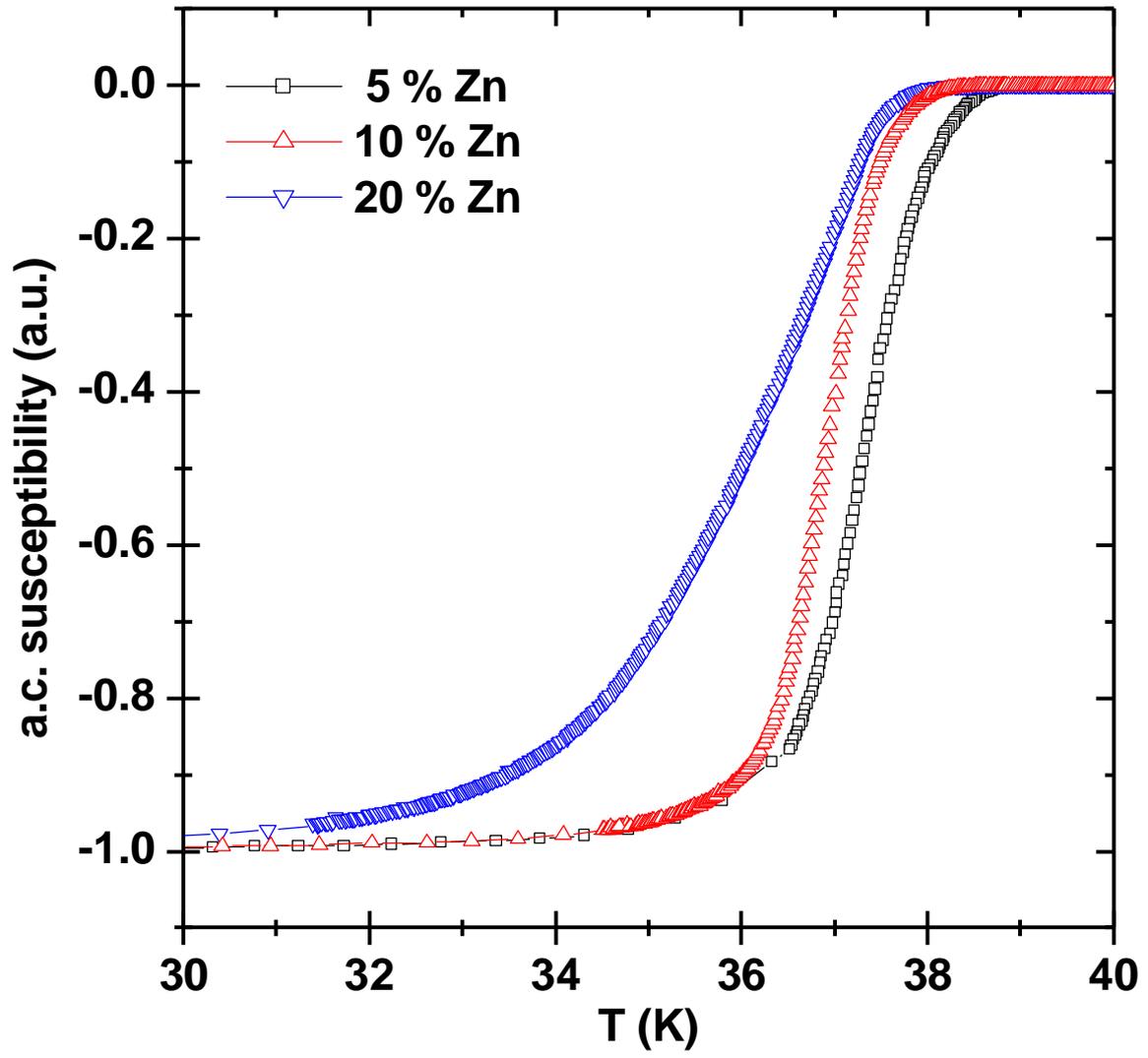

Fig. 8.

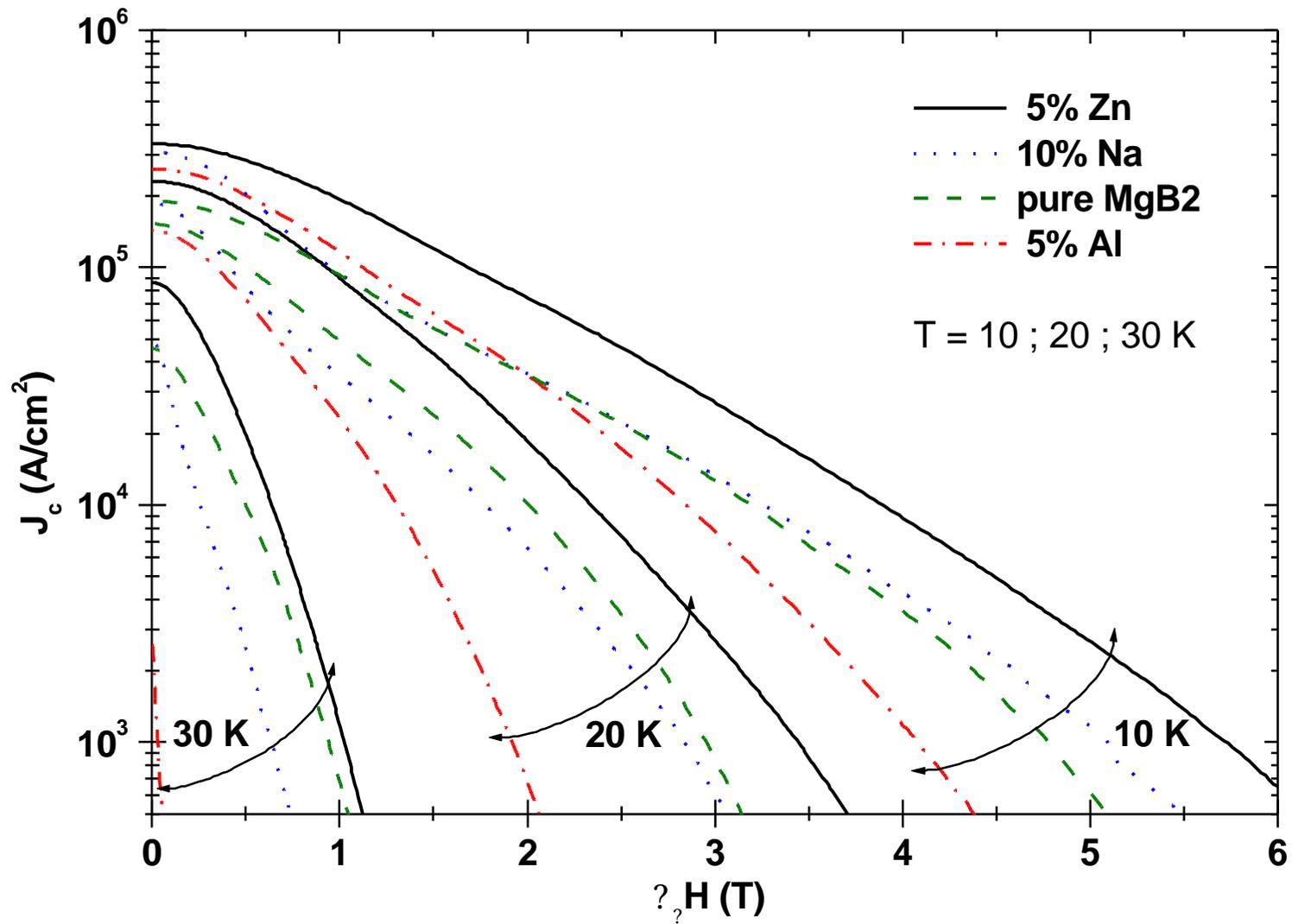

Fig. 9.

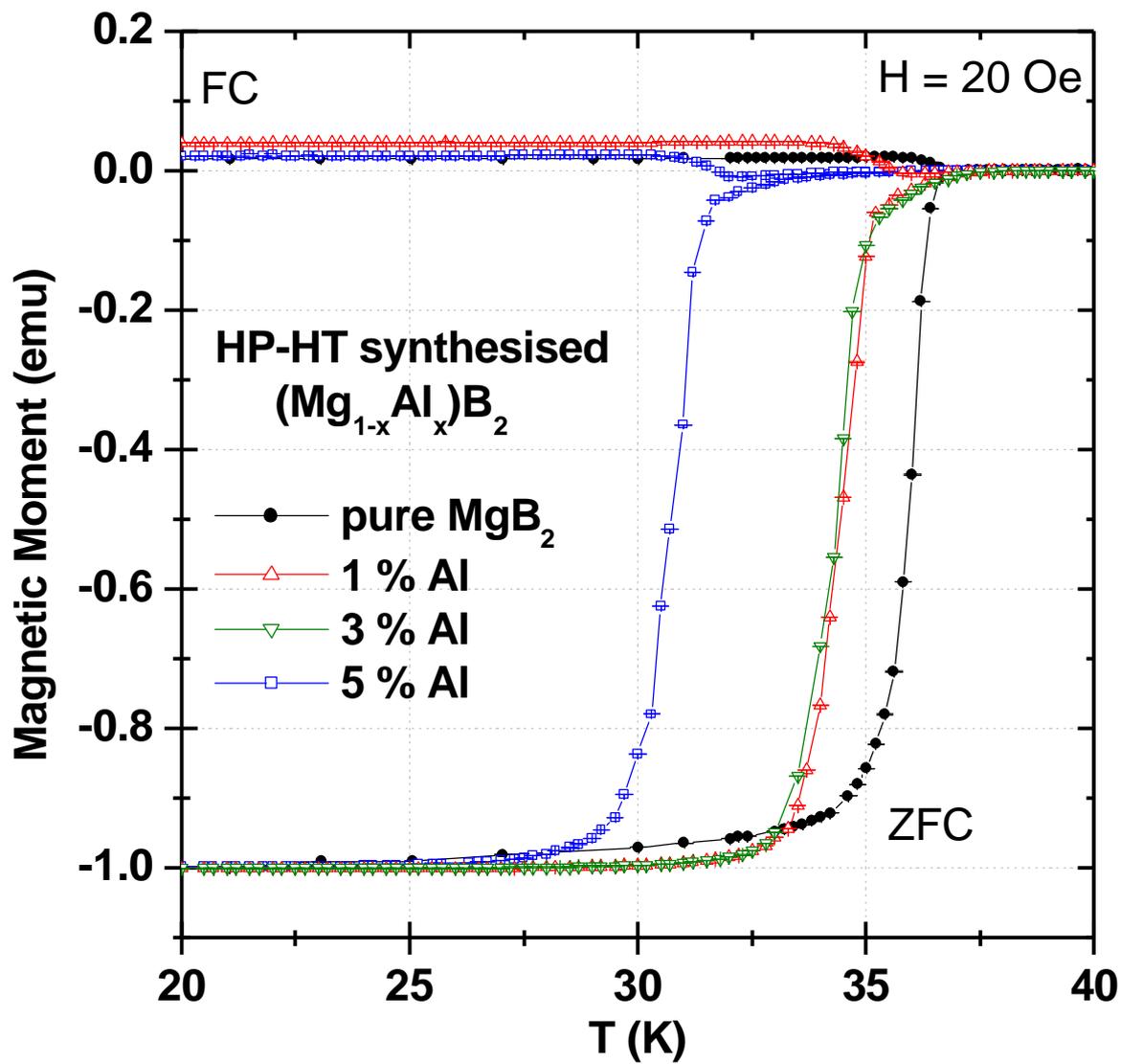

Fig. 10.

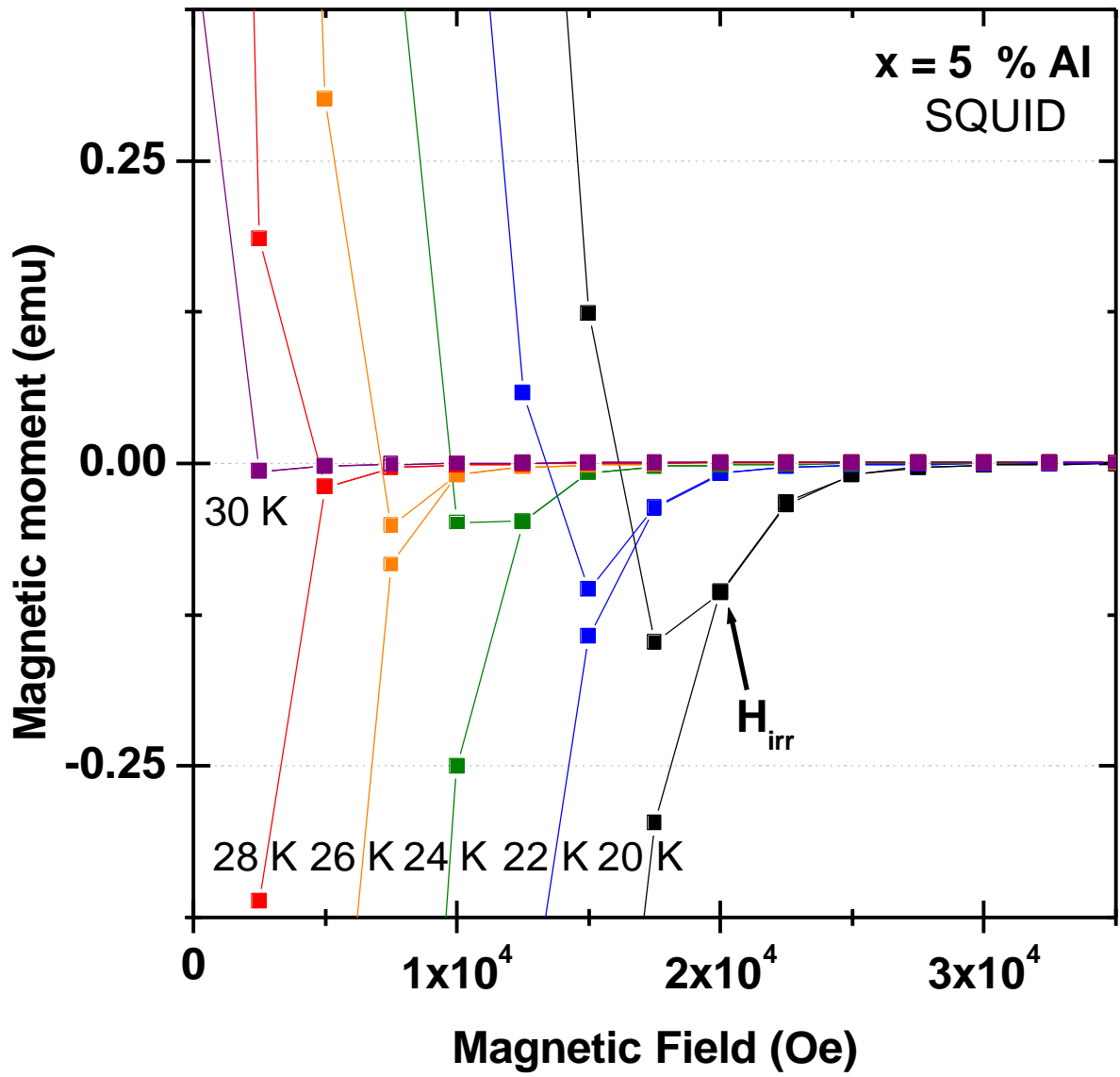

Fig. 11.

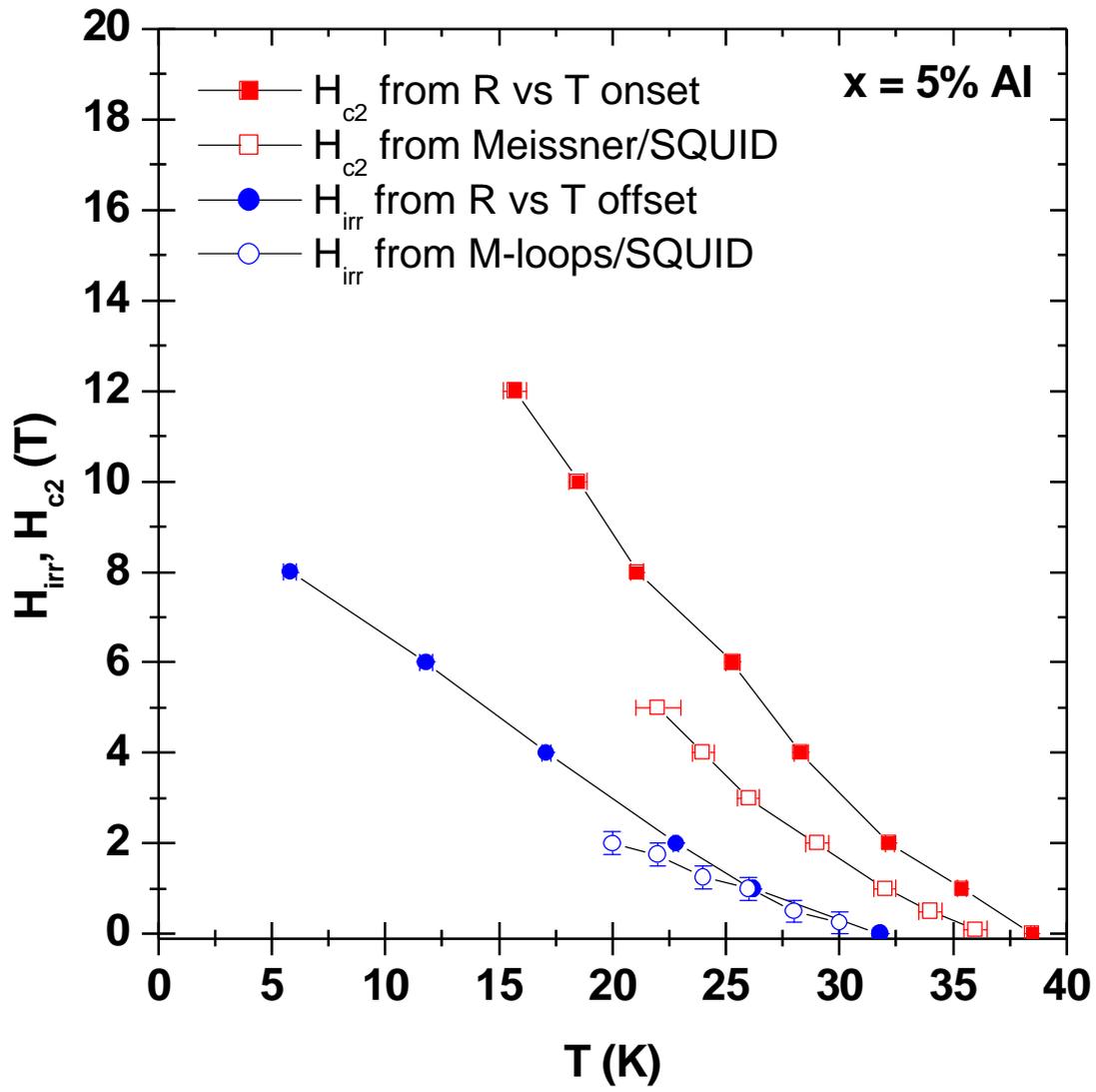

Fig. 12.